\documentclass[conference]{IEEEtran}
\IEEEoverridecommandlockouts

\usepackage[colorlinks, linkcolor=red, anchorcolor=blue,citecolor=blue,hyperfootnotes=true]{hyperref}
\usepackage[marginal]{footmisc}
\usepackage{tabularx} 
\usepackage{cite}
\usepackage{amsmath,amssymb,amsfonts}
\usepackage{algorithmic}
\usepackage{graphicx}
\usepackage{textcomp}
\usepackage{xcolor}
\usepackage{indentfirst}
\usepackage{float} 
\usepackage{subfigure} 
\usepackage{enumitem}
\usepackage{abstract}
\usepackage{blindtext}
\usepackage{nameref}

\newcounter{mylabelcounter}

\makeatletter
\newcommand{\labelText}[2]{%
#1\refstepcounter{mylabelcounter}%
\immediate\write\@auxout{%
  \string\newlabel{#2}{{1}{\thepage}{{\unexpanded{#1}}}{mylabelcounter.\number\value{mylabelcounter}}{}}%
}%
}
\makeatother

\begin{document}

\title{
  Literature review on vulnerability detection using NLP technology 
}

\author{\IEEEauthorblockN{1\textsuperscript{st} jiajie wu}
\IEEEauthorblockA{\textit{School of Computer} \\
\textit{Hangzhou University of Electronic Science and technology}\\
hangzhou,zhejiang, China \\
wujaijie@hdu.edu.cn}
}

\maketitle

\renewcommand{\abstractname}{}
\begin{onecolabstract}
\noindent{}Abstract:Vulnerability detection has always been the most important task in the field of software security. With the development of technology, in the face of massive source code, automated analysis and detection of vulnerabilities has become a current research hotspot. For special text files such as source code, using some of the hottest NLP technologies to build models and realize the automatic analysis and detection of source code has become one of the most anticipated studies in the field of vulnerability detection. This article does a brief survey of some recent new documents and technologies, such as CodeBERT, and summarizes the previous technologies.

\end{onecolabstract}


\section{Introduction}

In recent years, with the continuous maturity of software technology, more and more software has been developed by people. While people enjoy the convenience brought by software, they are also threatened by software vulnerabilities. It can be said that software vulnerabilities are one of the biggest problems that threaten the normal operation of software. For software users, the direct and indirect economic losses caused by software vulnerabilities worldwide have exceeded tens of billions of dollars. It is an indisputable fact that there are various vulnerabilities in most software. There are many types of software vulnerabilities, such as CVE-2015-8558\cite{akram2021sqvdt}, which are explained in detail on CVE\cite{CVE}. The longer the vulnerability exists, the easier it is to be exploited by hackers, and the greater the damage to the company or organization\cite{shin2010evaluating}, so the ability to automatically detect the vulnerability in the software within a certain time frame has become one of the hottest researches at the moment One.\\
\indent How can the automatic detection of vulnerabilities be more accurate? Deep learning technology gives us the possibility. With the continuous reform and development of deep learning technology in recent years, great progress has been made in the field of natural language processing (NLP). In particular, the series of models such as GPT\cite{radford2018improving} and BERT\cite{devlin2018bert} have taken NLP technology a big step forward. The source code is essentially a text in a special format. It is logically feasible to use NLP technology for code processing. In fact, in modern code intelligence, models such as CodeBERT\cite{feng2020codebert} have already been proposed by some scholars and some code-level tasks have been solved, and certain results have been achieved. These results show that the use of NLP technology to study automatic vulnerability detection (one of the code intelligence tasks) technology has a lot of room for development. We will do a detailed introduction in the chapter \ref{third_section}.\\
\indent The chapters are arranged as follows: Section \ref{sec_section} introduces the development of NLP, and section \ref{third_section} introduces the latest development of NLP technology in vulnerability detection. 

\section{The development of NLP technology}\label{sec_section}

\subsection{Natural Language Processing }
Natural language processing (NLP) is the use of computers to model human natural language in order to solve the application of natural language in some related problems. In NLP, the problems that need to be solved can be divided into two categories: 
\begin{itemize}
    \item 
    One is the natural language understanding (NLU) problem, including text classification\cite{kowsari2019text}, named entity recognition\cite{yadav2019survey,li2020survey}, relation extraction\cite{kumar2017survey}, reading comprehension, etc.\cite{dzendzik2021english,baradaran2020survey, zeng2020survey}; 
    \item 
    The second is natural language generation (NLG) problems, including machine translation\cite{chu2018survey,yang2020survey,dabre2020comprehensive}, text summary generation\cite{wang2020heterogeneous,mohd2020text}, automatic question and answer system\cite{sultana2020review,abbasiyantaeb2020text},Image caption generation\cite{katiyar2021comparative,parikh2020encoder,kalra2020survey} etc. 
\end{itemize}

When NLP researchers studied and solved these two types of problems, they found that the underlying problems that constitute these problems are basically the same, such as embeding expressions of vocabulary. Now researchers are more inclined to use a unified model for modeling (pre-training stage), and then adjust the model according to specific problems (fine-tuning stage). Research at this stage has made great progress. It is believed that in the near future, machines can truly understand human language and even understand human thinking. 

Since 1980s, traditional NLP has increasingly relied on statistics, probability and shallow learning (traditional machine learning)\cite{chowdhury2003natural}, such as naive Bayes, hidden Markov model, conditional random field, support Vector machines and K-proximity algorithms, etc., these algorithms are still widely used in NLP today. But with the development of deep learning (DL), people are paying more and more attention to how to use DL models to solve the problems in NLP\cite{otter2020survey}. 

\subsection{DL  in NLP}
The main goal of DL is to learn the deep neural network model\cite{alom2019state}. The neural network model is composed of neurons and the edges connected to them. Each neuron can input and output. The data inside the neuron can be nonlinearly transformed.\cite{schmidhuber2015deep}. According to the development of the timeline, we use the time point at which Transformer\cite{vaswani2017attention} is proposed as the segmentation point. The model method before its appearance is called the basic model method, and the later one is called the modern model method (or attention model method). We will introduce them separately below.
Basic model method introduction: 

\begin{enumerate}
    \item 
    Convolutional Neural Network (CNN)\cite{lecun1998gradient}: Due to the excellent abstract feature extraction ability of the convolution kernel, it has achieved great success in the field of computer vision (CV). In the field of NLP, CNN-based algorithms have also appeared one after another, such as\cite{kim2014convolutional, johnson2014effective, johnson2015semi, zhang2015sensitivity, nguyen2015relation}, etc. In the research related to vulnerability detection, some scholars have used CNN to mine vulnerabilities\cite{russell2018automated}, as shown in figure \ref{cnntoTEXT}. 
    \begin{figure}[htbp]
        \centering
        \setlength{\abovecaptionskip}{-0.5cm}
        \includegraphics[width=0.5\textwidth,height=40mm]{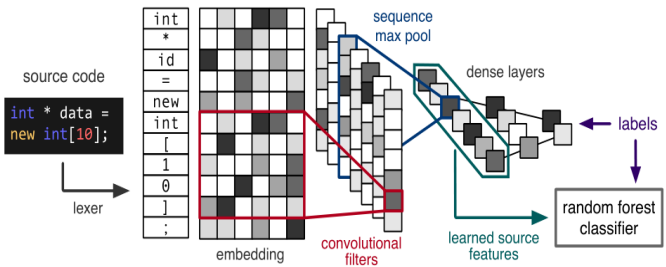}
        \caption{Using CNN to classify source code\cite{russell2018automated}}
        \label{cnntoTEXT}
    \end{figure}
    \vspace{-0.3cm}

    Although these models use CNN as a feature extractor to extract features from text data, because the feature dimensions of text data are not many, in text data, more attention is paid to the close connection between contexts, and the model is required to have a "memory" function , So CNN does not perform very impressively when processing tasks in NLP. But the latest research shows that with the development of multimodal technology\cite{baltruvsaitis2018multimodal, sulubacak2019multimodal, zhang2020multimodal}, in some code generation tasks, such as image generation instructions, the use of CNN-based models has achieved good results\cite{Katiyar_2020}. 

    \item 

    Recurrent Neural Network(RNN)\cite{kawato1987hierarchical}: One of the characteristics of RNN is its "memory". RNN can take serialized data as input or output serialized data. For serialized data such as text, using RNN for processing has a natural advantage. In the output of the RNN, the above sequence information of the current token can be included, which makes the RNN have a "memory" function. When processing the data in this article, people often use a two-way RNN, that is, to process the above and below information of the current token separately. Let the token contain the current context information at the same time, which is very important for the model to understand the meaning of the sentence. However, the model with RNN structure cannot be processed in parallel. Today, with massive data, it greatly reduces the development of RNN in engineering applications. In NLP, CNN and RNN are used to extract the character-level representation of words, as shown in Figure \ref{fig.cnnrnn}. 
    \begin{figure}[htbp]
        \centering
        \setlength{\abovecaptionskip}{-0.5cm}
        \includegraphics[width=0.5\textwidth,height=40mm]{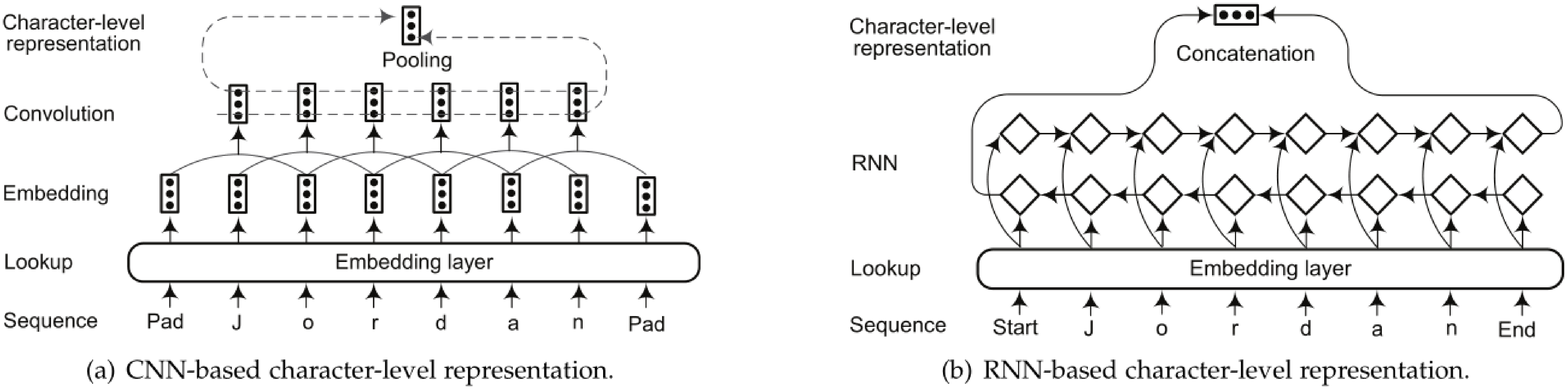}
        \caption{CNN \& RNN for extracting character-level representation for a word\cite{li2020survey}}
        \label{fig.cnnrnn}
    \end{figure}
    \vspace{-0.3cm}

    \item 

    Long Short-Term Memory Networks(LSTM)\cite{hochreiter1997long}: In addition to the structural limitations, RNN cannot capture long sequence text information due to the problem of vanishing gradient\cite{hochreiter1998vanishing}, so scholars modified RNN The LSTM model is proposed to solve the defect that RNN cannot process data in parallel. LSTM is one of the models with the strongest "memory" ability in NLP so far, and it is also one of the most widely used models. However, because LSTM has complex gating logic, it consumes a lot of space and time during training. Gated Recurrent Unit (GRU)\cite{cho2014properties} is a model that is similar in structure to LSTM but more lightweight, and its performance in training is not worse than LSTM. For the comparison between the three basic models of CNN, GRU, and LSTM in NLP applications, please refer to\cite{yin2017comparative}. Since LSTM is a one-way model, in order to obtain the context information of the token, people often superimpose the LSTM/GRU model in two directions to obtain a two-way LSTM model (Bi-LSTM)\cite{dai2019named}. In practical applications, the Bi-LSTM model is often used to extract the features of the sentence, and then the CRF algorithm is used to process the downstream tasks\cite{alzaidy2019bi}. 
    \item  
    Embedding\cite{bakarov2018survey,sivakumar2020review,limisiewicz2020syntax,ruder2019survey}: Embedding technology is a technology that can convert tokens into space vectors. The earliest embedding technology can be traced back to the distribution of words\cite{harris1954distributional}, which can represent a sequence of tokens in vector form as the input of the deep learning model. With the continuous development of technology, embedding technology can be divided into two types: 
    \begin{itemize}
        \item 
        One is the classic Non-Contextual embedding technology, which is also called contextual-independent embedding in some literatures, which refers to embedding independent of the context, such as Word2Vec\cite{mikolov2013efficient}, GloVe\cite{pennington2014glove} and other models. When embedding, the contextual semantic relationship of words in the sentence is not considered. To put it simply, these models only learn the mapping of words in the vector space. Each word is a fixed representation and cannot deal with the problem of word representation in the context of similar polysemous words. It is worth mentioning that in embedding technology, oov (out of vocabulary, oov) is often encountered. The common solution is to use substrings for further segmentation, such as BPE\cite{sennrich2015neural,wu2016googles}. For the semantic analysis performance of these classic models such as word2vec and GloVe, please refer to\cite{ccano2019word};     
        \item  
        The other is Contextual embedding technology, also called contextual-dependent embedding, such as the famous ELMo model\cite{peters2018deep,liu2020survey}, including the models such as BERT\cite{devlin2018bert} that appear later, the words embedding learned are all contextual embedding. Contextual embedding technology will comprehensively consider the context information in the sentence when learning the vectorized representation of words, and integrate the context information of a single token into the representation of the word. In this way, it is dealing with issues such as polysemous words, syntactic structure, and semantic roles. At the time, the words can be represented differently according to the current semantic environment. For a more detailed analysis and comparison of the two technologies, please refer to\cite{miaschi2020contextual}. 
    \end{itemize}
\end{enumerate}

Attention model method introduction:
\begin{enumerate}
    \item 
    Attention mechanism: The attention mechanism is an instinctive mechanism that imitates people when observing objects. In computers, the attention mechanism is essentially calculating the weight of a certain item, and finally all the items are weighted so that more information is contributed than the important items. The attention mechanism was first applied in machine translation\cite{bahdanau2014neural}. Due to its excellent performance, it was widely used in other NLP tasks. Now it has become very popular, and most of the NLP models have basically been integrated. The attention mechanism, especially in the encoder-decoder architecture, can be used alone in the encoder or decoder, or mixed, as shown in the figure \ref{fig.attentioninENDE}. In summary, the attention mechanism can be divided into 6 categories, as shown in the figure \ref{fig.attentionclassify}, of which the most common are self-attention and multi-dimensional attention. All models that apply the attention mechanism can be collectively referred to as Attention Model (AM). In addition to the application of AM in NLP, AM has also received extensive attention in the fields of Computer Vision (CV), Multi-Modal Tasks, Graph-based Systems and Recommender Systems (RS)\cite{chaudhari2019attentive}. 
    \begin{figure}[htbp]
        \centering
        \setlength{\abovecaptionskip}{-0.5cm}
        \includegraphics[width=0.5\textwidth,height=40mm]{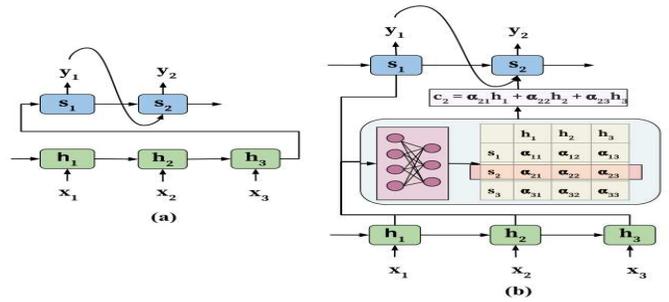}
        \caption{A comparison between the traditional encoder-decoder architecture (left) and the attention-based architecture (right)\cite{chaudhari2019attentive}}
        \label{fig.attentioninENDE}
    \end{figure}
    \vspace{-0.1cm}
    \begin{figure}[htbp]
        \centering
        \setlength{\abovecaptionskip}{-0.5cm}
        \includegraphics[width=0.5\textwidth,height=25mm]{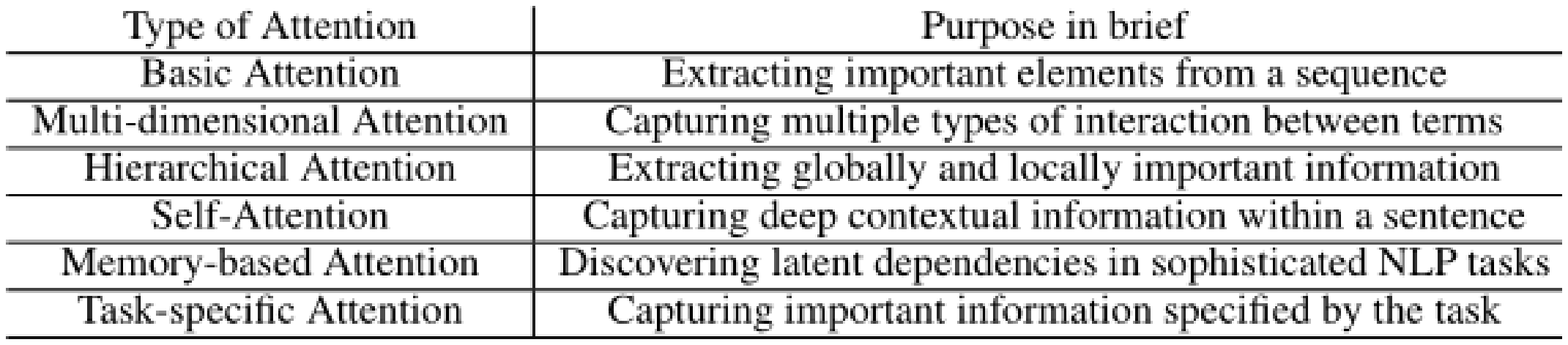}
        \caption{all attention types\cite{hu2019introductory}}
        \label{fig.attentionclassify}
    \end{figure}
    \vspace{0.1cm}
    \item 
    Transformer\cite{vaswani2017attention}: With the birth of the Transformer architecture, the architecture with the strongest feature extraction capabilities so far was born. In addition to the NLP field, Transformer has also made great progress in the CV field\cite{khan2021transformers}. The architecture of Transformer is shown in Figure \ref{Fig.transformerModel}. 

    As can be seen from the figure, Transformer integrates the attention mechanism. From the structural point of view, Transformer is a typical Encoder-Decode structure, and its general training process is as follows: 
     \begin{itemize}
         \item    
        On the encoder side, after the serialized token undergoes input embedding and positional embedding, the QKV matrix is generated using the three weight matrices of QKV, and then the attention matrix is obtained using the multi-head attention mechanism, and then passes through the conventional add\&norm, fully connected layer, etc. The whole process can be folded N times in total. In\cite{vaswani2017attention}, there are 6 folds;
         \item   
        On the decoder side, the process is roughly the same as that on the encoder side. The only difference is that a multi-head attention layer is added, that is, the second attention is performed. In this attention input, the v value uses the output of the encoder side. 
     \end{itemize}
    For specific details about Transformer and attention mechanism, please see\cite{ghojogh2020attention}. In\cite{tay2020efficient}, the author classified Transformers according to technology and main purpose. For the visualization research of Transformer, it is explained in this article\cite{bracsoveanu2020visualizing}, and we will not go into it here.
    \begin{figure}[htbp] 
        \centering 
        \setlength{\abovecaptionskip}{-0.5cm}
        \includegraphics[width=0.5\textwidth,height=60mm]{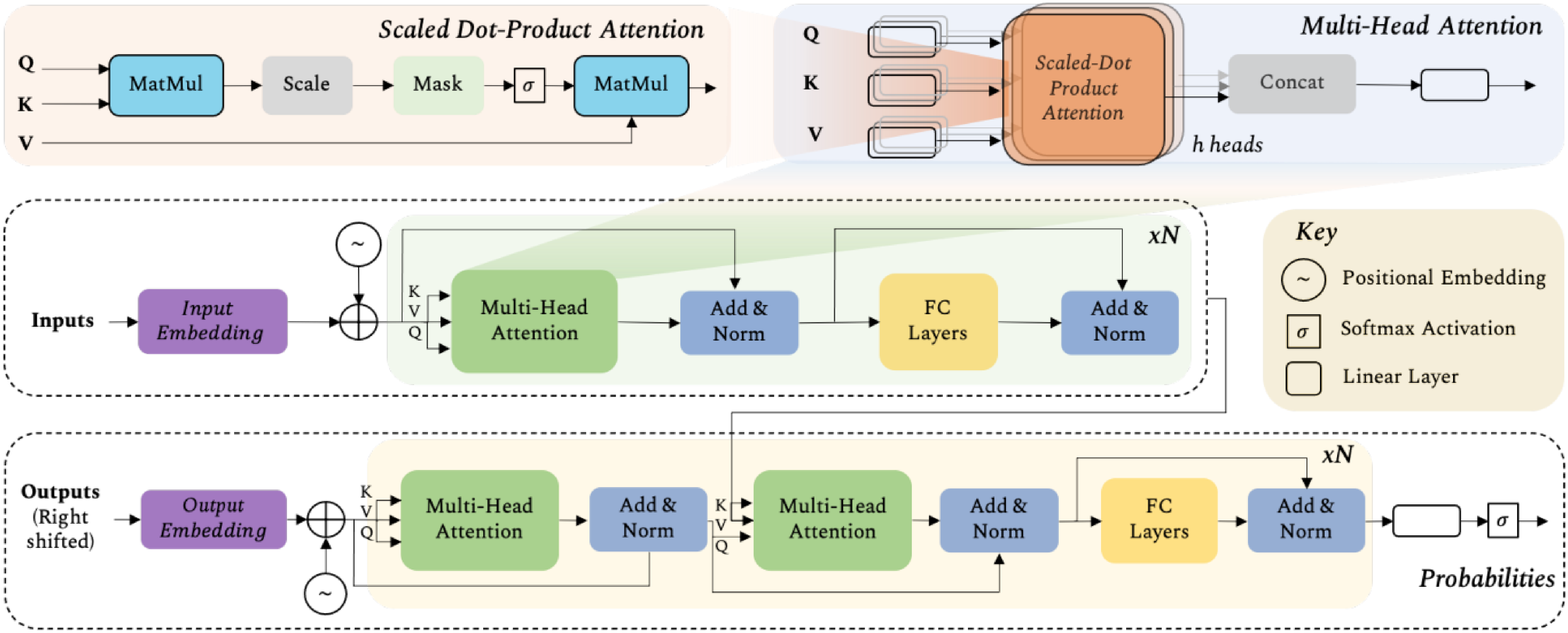} 
        \caption{Architecture of the Transformer Model\cite{khan2021transformers}} 
        \label{Fig.transformerModel} 
     \end{figure}
      \item 
    GPT\cite{radford2018improving}: The Generative Pre-trained Transformer (GPT) is a Transformer-based pre-training model developed by OpenAI. The purpose is to learn the dependency between sentences and words in long text. Over time, GPT has evolved from GPT-1 to GPT-3\cite{brown2020language}. The biggest difference between GPT-1 and BERT is that the GPT-1 model scans text from left to right, so token embedding can only consider the information before the current token, without considering the information below the token, while BERT uses a two-way model training . Therefore, GPT-1 only integrates the information above the token. For GPT-1's use of the information below the token, it is used as a new input to the model for training after prediction. GPT can realize unsupervised training. In GPT-3, unsupervised training of network text data is realized. The parameters in the model have reached 175 billion, which is about the number of GPT-2\cite{radford2019language} parameters (1.5 billion), GPT-3 can be said to be the largest and most advanced pre-training model so far.
      \item
    BERT\cite{devlin2018bert}: Bidirectional Encoder Representations from Transformers (BERT) is one of the best NLP models so far. BERT uses a two-way Transformer block for training, taking into account the context information contained in the word. After BERT, although many excellent models (such as XLNet\cite{yang2019xlnet}) have been proposed, the huge influence and excellent performance of BERT cannot be replaced by other models. The training process of BERT is shown in figure \ref{fig.bertpf}.
    
    \indent In BERT, Masked Language Modeling (MLM)\cite{taylor1953cloze} technology is used. This technique is a fill-in technique. When doing pre-training, it predicts the hidden information in the original text, and obtains the context embedding of the input token. The general process is that in the BERT input, approximately 15\% is randomly selected. The token of is masked, and then the BERT is pre-trained to predict the masked token. One disadvantage of this technique is that the masked token information will not be encoded into the context embedding. In the downstream task, the information deviation problem will occur due to the missing information of the previously masked word. The solution is to process the tokens selected to be masked at a random ratio of 8/1/1, that is, 80\% of the masked tokens continue to be masked, 10\% use the original token for training, and 10\% tokens are randomly replaced with other tokens. For a detailed summary of the application of the BERT model in NLP, please see\cite{koroteev2021bert}. 

       \begin{figure}[htbp]
           \centering
           \setlength{\abovecaptionskip}{-0.5cm}
           \includegraphics[width=0.5\textwidth,height=40mm]{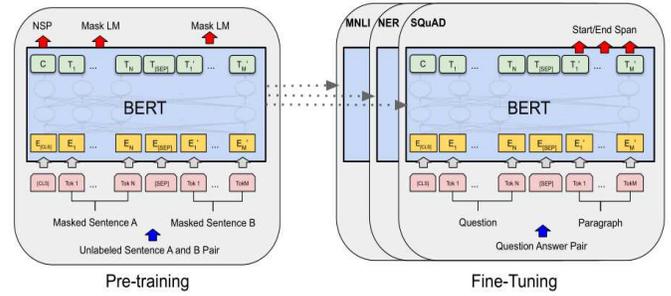}
           \caption{Overall pre-training and fine-tuning procedures for BERT\cite{devlin2018bert}}
           \label{fig.bertpf}
       \end{figure}
       \vspace{-0.3cm}

       \item 
    BART\cite{lewis2019bart}: BART is a denoising seq2seq algorithm, which can be said to be an extension of the BERT model. From an architectural point of view, it can be regarded as a "combination" of the BERT and GPT framework, using the encoder-decoder architecture , As shown in figure \ref{fig.bartat}.
       \begin{figure}[htbp]
           \centering
           \setlength{\abovecaptionskip}{-0.5cm}
           \includegraphics[width=0.5\textwidth,height=40mm]{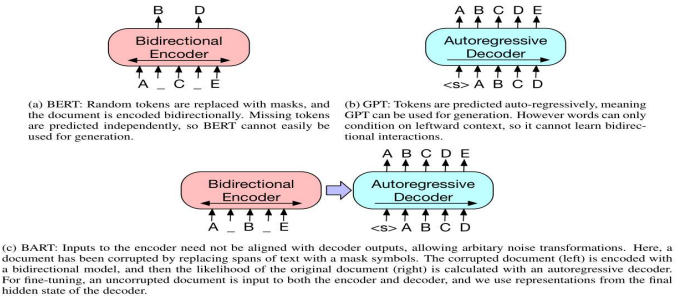}
           \caption{A schematic comparison of BART with BERT and GPT. \cite{lewis2019bart}}
           \label{fig.bartat}
       \end{figure}
       \vspace{-0.3cm}
    In the pre-training stage of BART, 5 noisy input transformation methods including Token Masking, Token Deletion, Text Infilling, Sentence Permutation, and Document Rotation are used. In the fine-tuning stage, the author trained four tasks: Sequence Classification, Token Classification, Sequence Generation, and Machine Translation. The results are shown in the figure \ref{fig.bartres}. As can be seen from the results in the figure, this extended model performs better than the BERT model on the data results. 
       \begin{figure}[htbp]
           \centering
           \setlength{\abovecaptionskip}{-0.5cm}
           \includegraphics[width=0.5\textwidth,height=40mm]{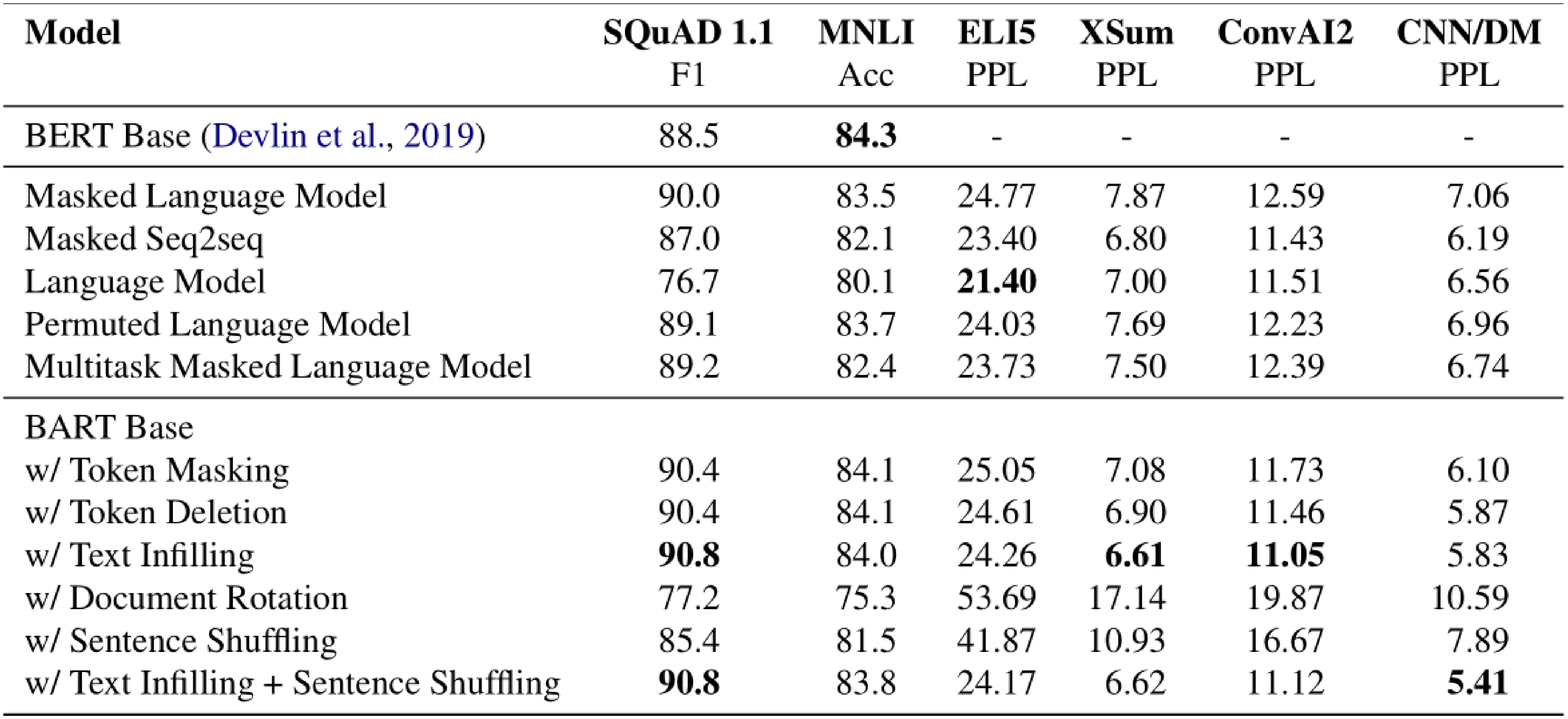}
           \caption{Comparison of pre-training objectives\cite{lewis2019bart}}
           \label{fig.bartres}
       \end{figure}
       \vspace{-0.3cm}
\end{enumerate}

\subsection{The pre-training model}


Currently, the mainstream research direction of NLP processing problems tends to be completed in two stages. The first stage is to build pre-trained models (PTM) based on context embedding. The second stage is based on Specific tasks fine-tuning the PTM. According to the classification in\cite{qiu2020pre}, PTM can be divided into three categories: serialization model, recursive model and self-attention model according to the model structure. Using the pre-training mechanism can improve the generalization performance of the model, allowing researchers or engineers to have more energy to deal with downstream specific tasks. It is worth mentioning that the bias problem in NLP will become prominent as the model becomes larger. For example, in GPT-3, the number of parameters has reached 175 billion. Although GPT-3 is by far the largest and most advanced NLP pre-training model, it also exhibits the most prejudiced\cite{garrido2021survey}. In addition, most of the pre-training models have a very large overhead (time, memory) during training. In some simple tasks, the effect of the context-independent embedding method is better than that of the context-dependent embedding\ cite{arora2020contextual}. This shows that there is no best model, only the most suitable model. To use a pre-trained model, there are usually two steps. The first step is to download the pre-trained model. You can use the third-party package transformer\cite{wolf2020transformers}. The second step is based on the specific downstream tasks. The model is fine-tuned. Generally, transfer learning\cite{zhuang2020comprehensive} is used to adjust the knowledge in the pre-training model to apply it to downstream tasks. There are many transfer learning methods in NLP, and the most widely used method is Domain Adaptation\cite{ramponi2020neural}. The article\cite{alyafeai2020survey} provides a more detailed classification of this. 

\section{vulnerability decetion using neural Networks}\label{third_section}

Vulnerability detection has always been the top priority in the field of software security. With the development of deep learning technology in CV, NLP and other fields, the use of deep learning methods to understand and detect vulnerabilities in the source code, thereby replacing manual detection methods, has become the focus and hotspot of current research\cite{lin2020software}. Although more and more detection methods have been proposed, the number of vulnerabilities reported on CVE\cite{CVE} and NVD\cite{NVD} is increasing day by day. The reason is that in addition to the large-scale increase in the number of software, another important reason is that root vulnerabilities are not easy to be detected, that is, if a root vulnerability is not detected, it will not help to repair other shallow vulnerabilities caused by it, and vice versa. , If the fundamental vulnerabilities are detected and fixed, other repetitive vulnerabilities will disappear. This requires vulnerability detection or mining tools to deeply understand the semantic information related to the vulnerability, so as to fundamentally detect the root vulnerability. To do this, deep NLP technology provides unlimited possibilities. 

\subsection{vulnerability introduction}
Software vulnerabilities are defined as follows\cite{ghaffarian2017vulnerability}, namely: \\
A software vulnerability is an instance of a flaw,caused by a mistake in the design, development, or configuration of software such that it can be exploited to violate some explicit or implicit security policy.\\
Vulnerability detection and analysis methods are divided into three types according to whether the detected code is executed or not: 
\begin{itemize}
    \item
    Static analysis: refers to the use of additional detection programs to detect programs that are suspected of vulnerabilities. During the analysis process, the detected program does not need to be executed, only the source code of the detected program is required; 
    \item 
    Dynamic analysis: refers to the execution environment that reproduces the software under test. Select the test cases required for the execution of the tested software, and then execute the tested program, monitor the program execution process and the variable change process, and find the loopholes in the execution in time; 
    \item 
    Hybrid dynamic and static analysis: As the name suggests, it refers to the use of dynamic analysis and static analysis together, but this does not essentially improve the accuracy of the analysis, because while focusing on the static and dynamic analysis points, it will also inherit the dynamic analysis and static analysis. The insufficiency. 
\end{itemize}
In this article, dynamic analysis techniques such as fuzzing testing or taint analysis\cite{wei2020smart, yulianton2020web, fell2017review, manes2019art, wang2020systematic} are not within the scope of this article. We only discuss static analysis techniques. According to whether the vulnerability detection technology uses the Transformer architecture, we artificially divide it into two categories, one is the detection model based on traditional DL technology, and the other is the NLP pre-training detection model based on the Transformer architecture (\ref{pretrainedModel}). 
Detection models based on traditional DL technology, such as LSTM/GRU/Bi-LSTM models, etc., when this type of model performs source code vulnerability detection, it is generally divided into two stages: 
\begin{enumerate}
    \item 
        The first stage is to segment the source code and extract the features in the source code. There are two ways to save the results after segmentation: 
        \begin{itemize}
            \item 
            One is based on the storage method of abstract syntax trees (AST). Use code attributes and use AST tree analysis tools to decompose the source code into the form of AST, and then perform vulnerability analysis in the AST tree\cite{yamaguchi2012generalized} or do other tasks, such as Alon uses path-AST (pAST) to express and complete the code The code completion task\cite{alon2019code2vec} has been added. 
            \item 
            One is the saving method based on the graph. Most of the graph segmentation results are saved as Code Property Graphs(CPG)\cite{yamaguchi2014modeling}. In CPG, AST, Control flow graph (CFG) and Program dependence graph (PDG) have been integrated together, and the extracted code feature information will be more, and the final vulnerability detection result will be relatively better, because in the CPG The vulnerability code provides more vulnerability information for the model, as shown in Figure \ref{cpgres}. Most of the literature now uses CPG to extract code features. For example, in\cite{wang2020combining}, CPG is called Augmented AST; in Devign\cite{zhou2019devign}, the sequence logic relationship between source codes (Natural Code Sequence, NSC) is actually another form of AST tree in CPG. It is worth mentioning that the data set used in Devign is widely used by many researchers, and \href{https://drive.google.com/file/d/1x6hoF7G-tSYxg8AFybggypLZgMGDNHfF/view?usp=sharing }{Data set open source}. There are many tools to generate CPG, you can directly use \href{https://github.com/ShiftLeftSecurity/joern}{Joern} or \href{https://github.com/mchalupa/dg}{DG}\cite{ chalupa2020dg} and other tools. The use of these tools is inextricably linked to \href{https://llvm.org/}{LLVM}. As for the AST tree generation tool, in the \url{https://github.com/Kolkir/code2seq} library, AST generation tools for programming languages such as Java, C++, C, C\# and python are provided. 
        \end{itemize}

        \begin{figure}[htbp]
            \centering
            \setlength{\abovecaptionskip}{-0.5cm}
            \includegraphics[width=0.5\textwidth,height=50mm]{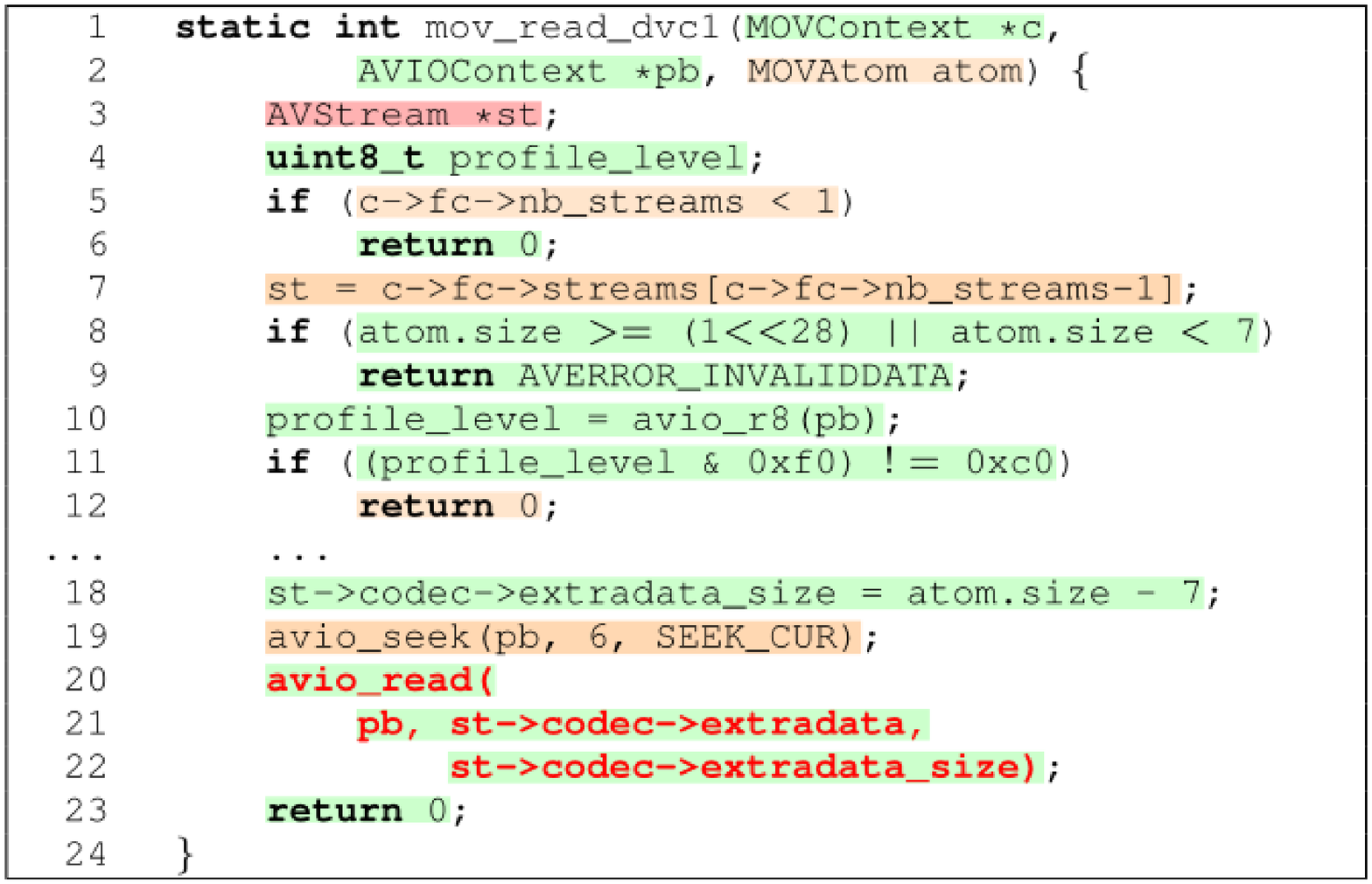}
            \caption{By graph splitting,the Red-shaded code elements are most contributing for vulnerability decetion\cite{chakraborty2020deep} } 
            \label{cpgres}
        \end{figure}
        \vspace{-0.1cm}
        
    \item 
        The second stage is modeling training. In this stage, the input is the output of the previous stage. According to whether the graph neural network (GNN) model is used or not, it can be divided into two categories: 
                
        \begin{itemize}
            \item   
            Use non-GNN model: Generally, the output in the first stage is in the form of a graph, so the graph needs to be encoded, converted into a vector, and then fed to the model. A series of work similar to SySeVR\cite{hu2019framework},Vuldeelocator\cite{Zou_2019,li2020vuldeelocator,zou2021interpreting,Liu2018heuristic}, the code is segmented at the token level, the semantic information inside the slice is relatively strong, and then Word2Vec\cite{mikolov2013efficient} is used for the slice code {mikolov2013efficient} Vectorized representation, which converts the slice code into a vector representation. In the modeling phase, these algorithms use LSTM or GRU and their variants Bi-LSTM, Bi-GRU and other models for modeling training, and the final results perform well on their respective artificialy synthesized data sets. But new research shows that\cite{chakraborty2020deep}, when tested with real data on the VulDeePecker\cite{Zou_2019} model, its accuracy is reduced to 11.12\%. This result is both unexpected and reasonable. Because the LSTM or Bi-LSTM model itself is not very sufficient in processing vulnerability information, the ability to extract relevant vulnerability information features is limited, that is, the generalization ability of the model is not strong, and the data in the real data set is unbalanced (not Vulnerability data is much more than vulnerability data), which causes the accuracy of the VulDeePecker model to be reduced by more than 50
            \item  
            Use GNN model: Since the source code is sliced in the first stage, it is generally saved as a graph. Therefore, continuing this logic, it becomes natural to use GNN for modeling training. In Devign\cite{zhou2019devign}, the gated graph neural networks (GGNN)\cite{beck2018graph,li2017gated} model is used for modeling training. The advantage is that the information in the entire graph structure can be fully considered, and there will be no adjacent junction information. Lost, more suitable for semantic graph structure representation in vulnerability detection tasks. When dealing with real data sets, the performance of existing detection models based on traditional DL technology is not very good. This is due to problems such as data imbalance and data duplication in real data sets. REVEAL\cite{chakraborty2020deep} can be used as a configurable vulnerability prediction tool, focusing on solving the problem of data imbalance in real data sets, and using representation learning to solve problems such as insufficient recognition of the vulnerability boundary by the model, as shown in Figure \ref{revealSP} The performance of the boundary between vulnerabilities and non-vulnerabilities under different models. In addition, wang\cite{wang2020combining} uses transfer learning in the model to deal with the problem of insufficient data. 
            \begin{figure}[htbp]
                \centering
                \setlength{\abovecaptionskip}{-0.5cm}
                \includegraphics[width=0.5\textwidth,height=25mm]{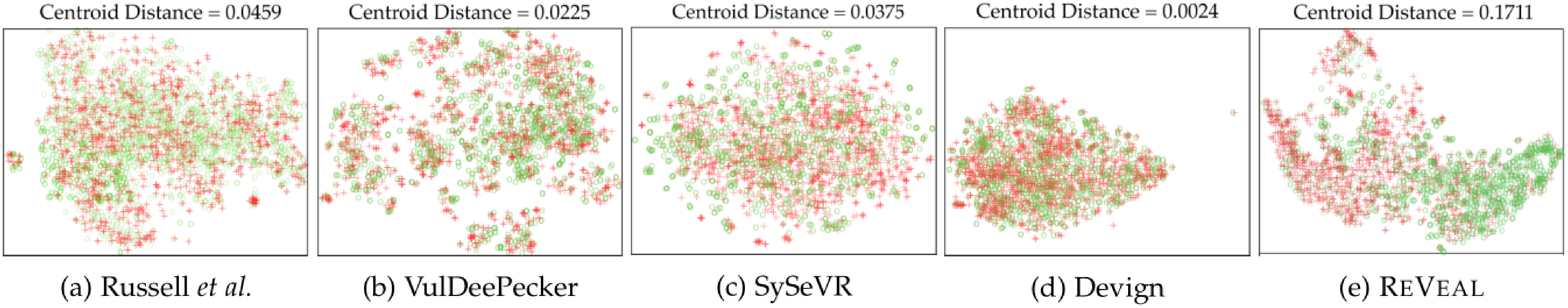}
                \caption{ t-SNE plots illustrating the separation between vulnerable (denoted by + ) and non-vulnerable (denoted by ◦ ) example\cite{chakraborty2020deep}}
                \label{revealSP}
            \end{figure}
            \vspace{-0.3cm}
        
        \end{itemize}

\end{enumerate}

\subsection{new era of  vulnerability detection }\label{pretrainedModel}

In the NLP field, the best models so far are models such as BERT\cite{devlin2018bert}, GPT\cite{radford2018improving} and their extended models. These models all use Transformer as the feature extractor. Since the code is also a special kind of text data, it is natural to think of using these excellent models such as BERT to do vulnerability detection. Listed below are some of the latest models that apply NLP technology to code intelligence (CI) tasks. These models have common characteristics, that is, the training process is divided into two stages, the first stage is pre-turning, and the second stage is fine-Tuning, and specific vulnerability detection tasks are generally completed in the fine-tuning stage. 
\begin{enumerate}
    \item 
    CodeBERT\cite{feng2020codebert}: CodeBERT is a model developed by Microsoft for code intelligence tasks. CodeBERT uses bimodal (bimodal)\cite{ramachandram2017deep,chen2020new,afridi2020multimodal} to train the model, where bimodal refers to natural language (NL) and programming language (PL), where NL refers to the program code Natural language annotations. In addition to the pre-training of the model using PL-NL dual-modality, CodeBERT also uses the pure code single-modality mode of 6 programming languages for training. In order to better adapt to this model, standard masked language modeling (MLM) and replaced token detection (RTD) methods are used for training, as shown in Figure \ref{codeBERT}. 
    \begin{figure}[htbp]
        \centering
        \setlength{\abovecaptionskip}{-0.5cm}
        \includegraphics[width=0.5\textwidth,height=40mm]{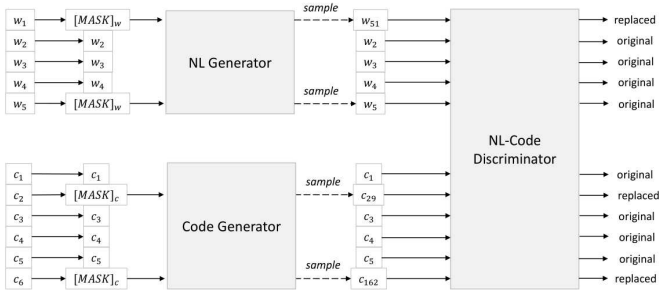}
        \caption{ CodeBERT training model\cite{feng2020codebert}}
        \label{codeBERT}
    \end{figure}
    It is worth mentioning that no model is a panacea. When using CodeBERT for code generation tasks, the code2seq\cite{alon2018code2seq} model does not perform as well. In code2seq, Alon uses the concept of path-context to extract more relevant semantic information from the code than CodeBERT, which uses source code for training. Later, in the extended version of CodeBERT GraphCodeBERT\cite{guo2020graphcodebert}, the internal structure of the code was considered. In the pre-training stage, the semantic-level structure of data flow was used to make the model more effective. In the four downstream tasks of code search, clone detection, code translation and code refinement, GraphCodeBERT achieved the best performance. 
    \item  
    CodeXGLUE\cite{lu2021codexglue}: In the code intelligence research, if a benchmark data set is provided, the research results will be more convincing. CodeXGLUE provides three types of model architectures: codeBERT, codeGPT, and code-encoder-decoder to help more researchers quickly solve problems in code intelligence. The problems in code intelligence that CodeXGLUE has implemented are specifically broken down into the following \labelText{four categories of sub-questions}{label:text}: 
    \begin{itemize} \label{codeTasks}
        \item
         code-code:clone detection, defect detection,cloze test,code completion, coderepair,code-to-code translation 
        \item
         text-code:natural language code search, text-to-code generation
        \item
         code-text:code summarization 
        \item 
        text-text:documentation translation
    \end{itemize}
    
    For detailed descriptions of these sub-problems, see\cite{lu2021codexglue}. In the latest pre-trained \href{https://huggingface.co/mrm8488/codebert-base-finetuned-detect-insecure-code}{CodeBERT} model, in the downstream task Insecure Code Detection, ACC has been increased to 65.3\% (previously 62.08\%) .
    \item 
    PLBART\cite{ahmad2021unified}: PLBART applies the BART\cite{lewis2019bart} framework to code intelligence, where PL refers to program language (PL). In PLBART, the noise reduction and self-encoding strategy in BART is continued, using token masking, token deletion, and token infilling three ways to add noise. In the fine-tuning stage, the author uses the four major tasks of Code Summarization, Code Generation, Code Translation, and Code Classification as the downstream tasks of PLBARK for fine-tuning. 
\end{enumerate}

Code Intelligence (CI) tasks refer to a series of tasks related to source code operations on the source code that are solved using artificial intelligence methods. Common code intelligence tasks are divided into \nameref{label:text}. This classification rule is the same as the classification rule of the problem in NLP, but the problem in NLP is oriented to the macro concept of "text", and the code is also A special kind of "text", we can regard vulnerability detection as a sub-task of code intelligence. The advantage of doing so is that more training samples and more generation pre-training models can be obtained. Applying some advanced NLP models to CI, using the powerful feature extraction capabilities of deep learning to extract relevant semantic information from the code, has gradually become a research hotspot. Research at this stage is mainly focused on the representation of vulnerability information. In other words, if deeper vulnerability information can be excavated, the ability to identify, judge and repair vulnerabilities will be greatly improved.

\section*{References}
\renewcommand\refname{}

\bibliographystyle{unsrt}

\bibliography{ref.bib}

\end{document}